\documentclass[sigconf]{acmart}

\AtBeginDocument{%
  \providecommand\BibTeX{{%
    \normalfont B\kern-0.5em{\scshape i\kern-0.25em b}\kern-0.8em\TeX}}}

\setcopyright{acmcopyright}
\copyrightyear{2022}
\acmYear{2022}
\acmDOI{XXXXXXX.XXXXXXX}

\acmConference[Websci '22]{}{26--29, June, 2022}{Barcelona, Spain}
\acmBooktitle{Websci '22: ACM Web Science Conference,
 26--29, June, 2022 Barcelona, Spain}
\acmPrice{15.00}
\acmISBN{978-1-4503-XXXX-X/18/06}

\usepackage{booktabs}
\usepackage{enumitem}
\usepackage{graphicx}
\usepackage{subfig}
\usepackage{hyperref}

\begin{document}

\title{Characterizing YouTube and BitChute Content and Mobilizers During U.S. Election Fraud Discussions on Twitter}


\author{Matthew C. Childs}
\email{mchilds3@vols.utk.edu}
\affiliation{%
  \institution{The University of Tennessee Knoxville}
  \city{Knoxville}
  \state{TN}
  \country{USA}
}

\author{Cody Buntain}
\email{cbuntain@umd.edu}
\affiliation{%
  \institution{University of Maryland}
  \city{College Park}
  \state{MD}
  \country{USA}
}

\author{Milo Z. Trujillo}
\email{milo.trujillo@uvm.edu}
\affiliation{%
  \institution{University of Vermont}
  \city{Burlington}
  \state{VT}
  \country{USA}
}

\author{Benjamin D. Horne}
\email{bhorne6@utk.edu}
\affiliation{%
  \institution{The University of Tennessee Knoxville}
  \city{Knoxville}
  \state{TN}
  \country{USA}
}

\begin{abstract}
In this study, we characterize the cross-platform mobilization of YouTube and BitChute videos on Twitter during the 2020 U.S. Election fraud discussions. Specifically, we extend the \texttt{VoterFraud2020} dataset~\cite{abilov2021voterfraud2020} to describe the prevalence of content supplied by both platforms, the \textit{mobilizers} of that content, the \textit{suppliers} of that content, and the content itself. We find that while BitChute videos promoting election fraud claims were linked to and engaged with in the Twitter discussion, they played a relatively small role compared to YouTube videos promoting fraud claims. This core finding points to the continued need for proactive, consistent, and collaborative content moderation solutions rather than the reactive and inconsistent solutions currently being used. Additionally, we find that cross-platform disinformation spread from video platforms was not prominently from bot accounts or political elites, but rather average Twitter users. This finding supports past work arguing that research on disinformation should move beyond a focus on bots and trolls to a focus on participatory disinformation spread.
\end{abstract}

\maketitle

\section{Introduction}
Social media has become a critical piece of our world's information infrastructure. Across social media, content is used to make decisions~\cite{cooley2019effect}, form social movements~\cite{jackson2020hashtagactivism}, and propagate cultures~\cite{zannettou2018origins}. Yet, the veracity and quality of that supplied content is not always equal. Disinformation, conspiracy theories, and hate speech have plagued social media platforms, forcing researchers, practitioners, and the platforms themselves to think deeply about methods to mitigate bad content's spread~\cite{bak2021combining, buntain2021youtube} and intervene in its consumption~\cite{horne2020tailoring}.

A key medium in this problem space is video, particularly social video platforms like YouTube~\cite{hussein2020measuring, papadamou2020just} and more recently, BitChute, an alternative to YouTube~\cite{trujillo2020bitchute, rauchfleisch2021deplatforming}. Video content is often linked to across multiple other social media, making social video platforms central in the proliferation of anti-social content online \cite{wilson2020cross, golovchenko2020cross}.

YouTube is the largest of such platforms, and, in the past, has been criticised for allowing disinforming content creators to flourish, and even make money from their content. For this reason and others, YouTube began making changes to how it recommends videos, what types of content can be hosted on the platform, and what types of content can be monetized \cite{youtube1, youtube2}. While there has been some evidence that these policy changes have been effective in reducing anti-social content online \cite{faddoul2020longitudinal, buntain2021youtube}, concerns about the effectiveness and the potential downstream impacts of these policy changes continue to be voiced. In general, the long-term impact of content moderation practices are still being debated and analyzed, with facets such as labor \cite{roberts2019behind}, regulation \cite{gorwa2019platform}, and post-moderation user behavior \cite{ali2021understanding} being discussed. 

One such concern is that with increased deplatforming and moderation on YouTube, content creators may be pushed from mainstream spaces to more extreme, un-moderated spaces like BitChute. Previous research has provided some evidence of this movement. For example, several of BitChute's most popular channels are content producers who were previously banned from YouTube, such as Alex Jones' Infowars \cite{trujillo2020bitchute}. Perhaps even more concerning is BitChute's ability to maintain deplatformed content's spread across other platforms. For example, BitChute allowed the viral COVID-19 conspiracy theory video, \textit{Plandemic}, to continue being shared after it was removed from mainstream platforms like Facebook, YouTube, and Twitter \cite{Kearney2020, Buntain2021}.

Even though there are growing concerns about YouTube and BitChute's roles online, relatively few studies have compared the cross-platform spread of their content. This lack of study is partially due to the difficulty of getting BitChute data, as it has no public API \cite{melabc}, and partially due to the need to examine the platforms within a single, comparable context. This paper begins to fill this gap through a descriptive analysis of YouTube and BitChute content shared on Twitter during the discussions of election fraud in the 2020 U.S. Presidential Election. Specifically, by combining two publicly available datasets and utilizing this single discussion context, we explore four questions:
\begin{itemize}
    \item \textbf{RQ1:} How prevalent was content supplied by BitChute and YouTube on Twitter?
    \item \textbf{RQ2:} How were the mobilizers of BitChute content different than the mobilizers of YouTube content?
    \item \textbf{RQ3:} How was the content supplied by BitChute different than the content supplied by YouTube?
    \item \textbf{RQ4:} How were the content suppliers on BitChute related to the content suppliers on YouTube?
\end{itemize}

We found that despite the growing concern about BitChute's role online, YouTube is still more prevalent in the spread of disinformation. Specifically, during the 2020 election fraud discussions on Twitter, YouTube videos promoting election fraud claims were linked to approximately 28 times more than BitChute videos promoting election fraud claims. Tweets linking to YouTube videos promoting fraud claims also received more engagement than tweets linking to BitChute videos promoting fraud claims. Specifically, we found that YouTube linked tweets from promoters received 14.10 retweets on average, while BitChute linked tweets from promoters received only 7.56 retweets on average. Furthermore, the majority of the video channels that supplied this content were unique to YouTube, rather than appearing on both platforms.

This core finding is not to say that BitChute and other alt-tech platforms should not be studied. While BitChute content was linked to magnitudes less than content from YouTube, that content was nearly all promoting election fraud claims and often mixed together with multiple other conspiracy theories. More precisely, tweets with links to BitChute were 302 times more likely to be promotions of election fraud claims rather than detractors from, while tweets with links to YouTube were only 21 times more likely to be promotions of election fraud claims rather than detractors from. 

Overall, these results point to the need for continued efforts in building and researching \textit{proactive} moderation solutions rather than the current \textit{retroactive} solutions. Perhaps even more importantly, these solutions should be collaborative and consistent across major platforms, as the content policies on one platform clearly impact the content spread on other platforms. This paper adds further evidence to the already growing body of literature about this need. 

\begin{figure}[ht]
    \centering
    \includegraphics[width=0.46\textwidth]{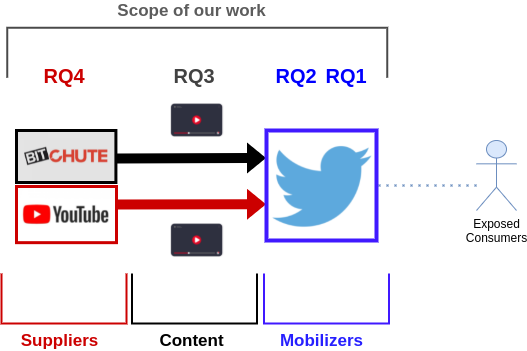}
    \caption{We view this work through a `Supplier-Mobilizer' framework, in which we describe: the \textit{suppliers} of content from YouTube and BitChute (channels), the content itself (video titles), and the \textit{mobilizers} of that content on Twitter (Twitter accounts linking to videos).}
    \label{fig:flow}
\end{figure}

\section{Related Work} 

\subsection{Claims of election fraud in the 2020 U.S. Presidential Election.} 
Before introducing the scholarship that motivated our research questions, we first want to introduce the context of our data. During the 2020 U.S. Presidential Election, Democratic Party candidate Joe Biden was elected President. Both before and after the election, right-leaning politicians and influencers made efforts to delegitimatize the election process and its outcome \cite{election_integrity}. False claims, crafted narratives, and conspiracy theories centered around every step in the election process, from claims of citizens voting multiple times using mail-in ballots to hacked voting machines changing votes. Ultimately, this multi-faceted set of claims built up a narrative that the election was stolen from then incumbent Donald Trump, inspired the online \#StopTheSteal movement, and eventually motivated the attempted insurrection on January 6th, 2021. 

Important in motivating our study of cross-platform content, a report by The Election Integrity Partnership described the production and dissemination of misinformation during this event as ``multidirectional and participatory" as well as ``cross-platform \cite{election_integrity}.'' Fact-checkers and platform moderators had inconsistent, unclear, and limited success curbing amplification of misinformation, which fell into three categories: bottom-up (claims made by individuals on social media), top-down (claims made by political influencers both on and off social media), and cross-platform (claims made on and spread across multiple social media platforms) \cite{election_integrity}. Hence, these complications allowed for harmful content to live and spread on both mainstream platforms and alt-tech platforms, and we argue that the dynamics between these platforms warrants further study.


\subsection{Content moderation on YouTube}
In early 2019, YouTube made changes to its recommendation algorithm to reduce the spread of content that ``comes close to-but doesn't quite cross the line of- violating [their] Community Guidelines \cite{youtube1}.'' This content included potentially misinforming videos such as "phony miracle cure[s]'' and 9/11 conspiracy theories \cite{youtube1}. In mid-2019, YouTube made further changes to content moderation, this time focused on removing hateful content (e.g. content that ``glorif[ies] Nazi ideology'') and tightening restrictions on who can use the monetization features on the platform (a similar restriction on monetization features happened in 2017) \cite{youtube2}. 

To some extent, there is evidence that these policy changes have had an impact both on YouTube and on the greater ecosystem. For example, \citet{faddoul2020longitudinal} indicates that recommendations to anti-social content on YouTube have been reduced and \citet{buntain2021youtube} provide evidence that when YouTube began de-recommending ``potentially harmful'' videos, the spread of conspiracy-focused videos on both Twitter and Reddit decreased. On the other hand, most content moderation practices, by YouTube and other major platforms, have been characterized as `reactionary' rather than preventive, meaning that only when bad content gains high attention is it dealt with. Hence, while we have seen a reduction in known anti-social content on YouTube, it remains an open question as to how well YouTube's content moderation policies, and others, can keep up with emerging disinformation events and campaigns. Maybe more importantly, it is an open question of how well content moderation policies across multiple platforms can work together during these events. The report by The Election Integrity Partnership on misinformation during 2020 election indicates as much, stating: ``Platforms took action against policy violations... [h]owever, moderation efforts were applied inconsistently on and across platforms, and policy language and updates were often unclear \cite{election_integrity}."

No matter the effectiveness of these content moderation policies, the simple act of moderating (suspending accounts, removing content, etc.) may have an unwanted side effect. As more anti-social content and content producers have been removed from and de-recommended by major platforms like YouTube, self-proclaimed alternatives to ``big-tech'' platforms have emerged (often called \textit{alt-tech} \cite{Wilson2021}). In particular, the platform BitChute emerged as an alternative to YouTube in 2017. BitChute has been characterized as ``low-rent YouTube clone that carries an array of hate-fueled material \cite{splc1}'', and has been growing \cite{trujillo2020bitchute}. While very little work has explored BitChute's relationship with YouTube explicitly, there is evidence that content producers overlap between the platforms \cite{trujillo2020bitchute} and that when YouTube removes bad content, that content can transition to BitChute, allowing the bad contents continued spread in the greater ecosystem \cite{buntain2021youtube}. From the previous work on disinformation during the 2020 U.S. election, we know of at least one instance where video content was removed from mainstream platforms and republished on BitChute \cite{election_integrity}. 

The reduction of anti-social content on YouTube and the growth of the YouTube alternative, BitChute, leads us to our first research question: \textbf{RQ1:} How prevalent was content supplied by BitChute and YouTube in election fraud discussions on Twitter? While past work and media coverage suggest that BitChute and other alt-tech are playing an increasing role in the anti-social information space, there lacks a grounded comparison between the role of YouTube and the role of BitChute. The online discussion around election fraud during the 2020 election provides us with a single context to examine this prevalence in. 


\subsection{Cross-platform spread of disinformation}
In general, we know that content is often mobilized across other social platforms, and the cross-posting of video content is particularly prevalent. For example, \citet{wilson2020cross} noted that anti-White Helmet operations used YouTube as a resource in their Twitter campaign. Similarly, \citet{golovchenko2020cross} showed that the Internet Research Agency (IRA) leveraged content from YouTube in their 2016 propaganda campaign on Twitter. These works demonstrate that content supplied by YouTube content creators is not just engaged with on the platform, but can be disseminated across other social platforms, sometimes purposely and maliciously so. 

In these cases of cross-posted video content, the mobilizers of that content can be bots, sockpuppets, influential elites, or everyday information consumers. This leads us to our second research question, \textbf{RQ2:} How were the mobilizers of BitChute content different than the mobilizers of YouTube content? The make-up of the accounts that shared tweets linked to videos can further characterize the roles of YouTube and BitChute in the ecosystem. For instance, if regular Twitter users only link to YouTube content, while bot accounts or political elites link to BitChute content, the roles of BitChute and YouTube as the suppliers of that content are different. Our second research question explores this idea.

\subsection{Relationship between content and channels on YouTube and BitChute}
Lastly, given the size of YouTube compared to BitChute, it is likely that YouTube supplied more content to the discussion on Twitter than BitChute. However, even if the sheer number of YouTube videos linked to during discussion is much larger than the number of videos linked to on BitChute, the content of those links could vary drastically in veracity and topic. Alternatively, given the known overlap in YouTube and BitChute content creators \cite{trujillo2020bitchute}, the content supplied by both platforms may be indistinguishable. This leads us to our third and fourth research questions: \textbf{RQ3:} How was the content supplied by BitChute different than the content supplied by YouTube? and \textbf{RQ4:} How were the content suppliers (creators) on BitChute related to the content suppliers (creators) on YouTube?  Overall, very few works have attempted to characterize and compare the content across big-tech and alt-tech platforms. The results from these two questions add to this limited literature.

\section{Data}
To answer our research questions, we combine two publicly-available datasets: the \texttt{VoterFraud2020} dataset~\cite{abilov2021voterfraud2020} and the \texttt{MeLa-BitChute} dataset \cite{melabc}.

\subsection{The VoterFraud2020 Dataset}
The \texttt{VoterFraud2020} dataset contains 7.6M tweets and 25.6M retweets related to election fraud claims during the 2020 U.S. Presidential Election \cite{abilov2021voterfraud2020}. The data was collected using the Twitter streaming API over a manually curated set of hashtags and keywords related to election fraud claims. The data was collected between October 23rd, 2020 and December 16th, 2020. Based on \citet{abilov2021voterfraud2020}'s analysis, the dataset covers roughly 60\% of the content on Twitter that used the tracked keywords.

In addition to data from the Twitter API, the dataset is enhanced with a variety of rich metadata. Important to our analysis in this paper are two pieces of metadata. First, the authors collected YouTube metadata for videos linked to in the tweets. More precisely, given that YouTube links made-up over 12\% of the external URLs in the dataset, the authors retrieved metadata for videos still available on January 1st, 2021 from the YouTube API. Of the 13,611 unique video URLs, 12,003 of them were still available for metadata collection. The YouTube metadata includes video title, channel, and video description.

Second, the authors labeled Twitter users as ``promoters'' of or ``detractors'' from election fraud claims. To do this labeling, the authors created a retweet network where nodes represent Twitter users and directed edges represent if one Twitter user retweets another. In this network, edges are weighted by the number of times a target node retweets a source node. With this weighted, directed network, the authors detected communities using the Infomap community detection algorithm \cite{bohlin2014community} and qualitatively determined the type of content being retweeted in each community. These communities are then used to label users as promoters of or detractors from election fraud claims. For more details on this network formation, we refer you to the \texttt{VoterFraud2020} paper and website\footnote{\url{https://voterfraud2020.io/}}.

\subsection{The MeLa-BitChute Dataset}
To enhance our extracted data from the \texttt{VoterFraud2020} dataset, we extract metadata for all linked to BitChute videos using the \texttt{MeLa-BitChute} dataset. The \texttt{MeLa-BitChute} dataset contains data from 3M videos published by 61K channels between June 2019 and December 2021 on BitChute \cite{melabc}. Since BitChute does not have a public API for data collection, this data was collected using a custom built scraper and contains metadata on nearly every video published on the platform during that time. For this paper, we match BitChute URLs from the \texttt{VoterFraud2020} dataset and extract metadata from the BitChute videos to match the metadata already extracted from the YouTube videos: video title, channel, and video description. 

In total, we analyze 83K tweets with links to YouTube or BitChute from 37K Twitter accounts, and from those links, we map and analyze 13K unique videos from 5084 YouTube channels and 342 BitChute channels.

\section{Methods and Results}

\subsection{RQ1: How prevalent was content supplied by BitChute and YouTube on Twitter?}

\begin{figure}[h]
    \centering
    \subfloat[\centering Number of Tweets containing a link to each social media platform (Log-scale).]{{\includegraphics[width=0.38\textwidth]{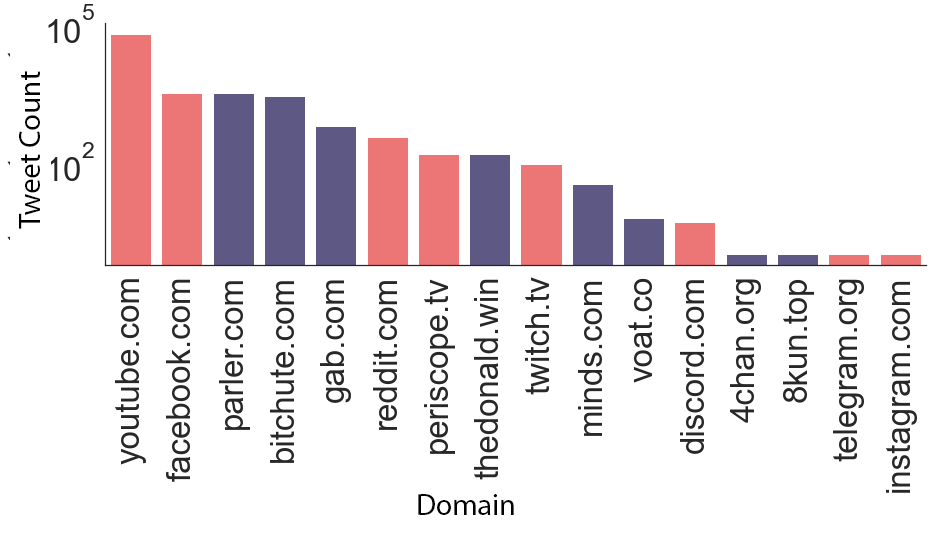}}}%
    \qquad
    \subfloat[\centering Number of interactions to tweets containing a link to each social media platform (Log-scale).]{{\includegraphics[width=0.38\textwidth]{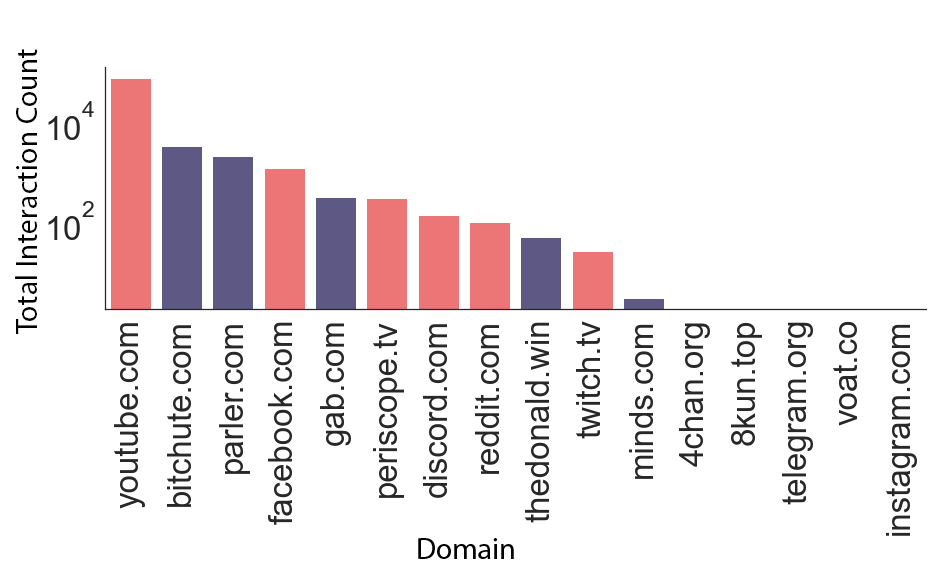}}}\\
    \includegraphics[width=0.35\textwidth]{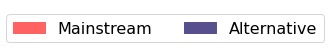}
    \caption{Prevalence of content from other social media platforms on Twitter during voter fraud discussion. Links to BitChute were tweeted approximately as much as links to much larger platforms like Facebook, and those tweets had slightly more interactions than tweets linking to Facebook. Note, we are counting all tweets with URLs to these platforms, rather than unique URLs.}%
    \label{fig:prevelence}%
\end{figure}

First, we want to know how prevalent cross-posting content supplied by BitChute and YouTube was in the discussion on Twitter compared to other social media platforms. In order to answer this question, we extract all tweets from the \texttt{VoterFraud2020} dataset which contained a URL with the domain of one of 16 predefined social media platforms. This predefined list included both established platforms and emerging fringe platforms. See Figure~\ref{fig:prevelence} for this list. Note, we are only looking at URLs to these platforms, rather than all URLs in the dataset. As described in \citet{abilov2021voterfraud2020}'s work, URLs to news outlets, both fringe and mainstream were frequently shared in the discussion.   

As expected, we found that tweets with links to content from YouTube occurred magnitudes more often than links to content from any other platform. This finding is also made clear in Figure \ref{fig:prevelence}. However, importantly, links to content from several fringe platforms also occurred frequently in the discussion. Namely, both tweets with links to Parler content and links to BitChute content occurred approximately as much as tweets linking to content from Facebook. Content from each of the three platforms (Facebook, Parler, and BitChute) tied for second in magnitude. Alone in a distant third place was Gab, followed by Reddit in fourth place. 

Similarly, when examining tweet interactions (retweets and quote tweets) we see fringe platforms rank near the top. Again, tweets with links to YouTube content ranked magnitudes above the rest. More interestingly, we see tweets with links to BitChute content were interacted with slightly more than tweets with links to Parler content and tweets with links to Facebook content. Given the increasing popularity of video content, noted in several studies \cite{perrin2019share}, this result may seem unsurprising, but also provides some evidence that BitChute has an emerging role in the ecosystem.  

Overall, while YouTube content was by far the most prevalent mobilized content in the Twitter discussion, content from fringe, low-moderation platforms like BitChute and Parler were as prevalent as content from Facebook, and much more prevalent than content from other mainstream platforms like Reddit. 

\begin{figure*}[ht!]
    \centering
    \includegraphics[width=0.85\textwidth]{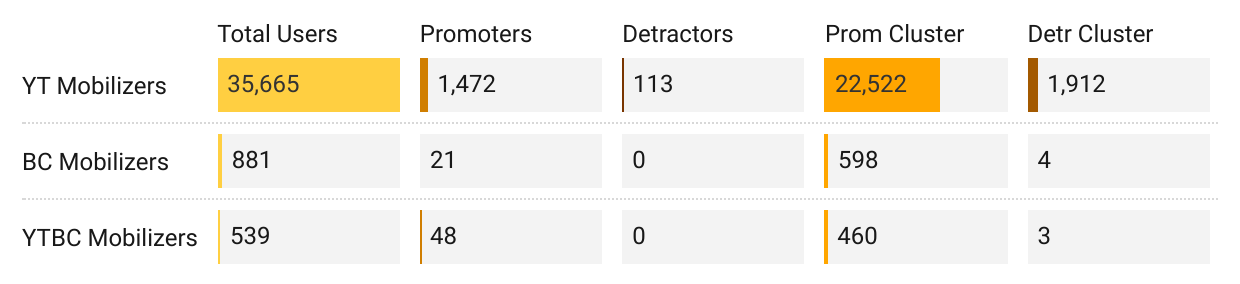}
    \captionof{table}[]{Number of promoters and detractors in each user group. Using the labels from \cite{abilov2021voterfraud2020}, we count the number of promoters of and detractors from election fraud claims in each user group. Since only the most central 10K users were labeled as Promoters or Detractors in the VoterFraud2020 dataset, we also extend this count to all of the users in the promoter clusters and detractor clusters as described in \cite{abilov2021voterfraud2020}. \label{tbl:mobgroups}}
\end{figure*}

\subsection{RQ2: How were the mobilizers of BitChute content different than the mobilizers of YouTube content?}
Now that we have established that both YouTube and BitChute played some role in the discussion on Twitter (although of vastly different magnitudes), we shift our attention to the mobilizers of the content from the two platforms. Explicitly, we define \textit{mobilizers} as the Twitter accounts who tweet or retweet links to content from YouTube and/or BitChute. Again, our goal is to understand if those accounts who shared BitChute links instead of YouTube links were significantly different, potentially indicating different audiences between the platforms or different types of elite mobilizers, such as far right-leaning influencers or other strategic information campaigns. 

To explore the similarities and differences, we extracted three groups of Twitter users from the data: \textit{YouTube (YT) Mobilizers}, \textit{BitChute (BC) Mobilizers}, and users who shared links to both platforms (\textit{YTBC Mobilizers}). With these user groups created, we extracted all tweets made by those users.

We then explore four facets about these groups:
\begin{itemize}
    \item Were mobilizers promoters or detractors of election fraud claims? 
    \item How many mobilizers were eventually suspended by Twitter?
    \item Were mobilizers verified on Twitter or bot accounts?
    \item Were BC mobilizers sharing BitChute content before the election?
\end{itemize}

\subsubsection{\textbf{Were mobilizers promoters of or detractors from election fraud claims?}} \label{sec:promdetr}
In order to answer this question, we utilize the promoter and detractor labels provided in the \texttt{VoterFraud2020} dataset. In Table~\ref{tbl:mobgroups}, we show two breakdowns of the user groups. First, we count the number of mobilizers who are in community 0, determined to be the detractor cluster in the \texttt{VoterFraud2020} dataset, and mobilizers who are in communities 1, 2, 3, and 4, determined as promoter clusters in the \texttt{VoterFraud2020} dataset. Second, we use the more strictly defined label of promoters and detractors from the dataset, which were the 10K most central nodes within the promoter and detractor clusters \cite{abilov2021voterfraud2020}. Note, the number of users in the promoter clusters and the detractor cluster do not add up to the total number of users in each mobilizer group. This difference is due to two reasons: (1) Nodes were randomly sampled in the original network creation and (2) All communities that contained fewer than 1\% of the nodes were excluded from the community analysis in the original dataset \cite{abilov2021voterfraud2020}. Hence, some mobilizers did not fall into one of the 5 largest communities or were missed in the random sampling. 

Unsurprisingly, we found that all users who shared a BitChute link were labeled as promoters of election fraud claims. Only seven BC and YTBC mobilizers were in the detractor community, while none of those users were labeled as detractors when using the more strictly defined metric. A qualitative review of the tweets from these seven users showed that the content was clearly aimed to promote claims of voter fraud. So, we suspect these users fell on the peripheral of the detractor community.

Perhaps more surprisingly, we found that users who shared YouTube links were also highly likely to be promoters of election fraud claims, with only 113 labeled as detractors and 1472 labeled as promoters. When expanding to the broader community label, all three sets of mobilizers still fell overwhelming in the promoter camp, with just slightly more YT mobilizers detracting from the claims, proportionally. In other words, if a Twitter user shared a link to video content, they had approximately a 92\% chance of being a promoter of election fraud claims (using the broader community labels). 

The engagement with tweets by promoters and detractors tells a similar story. Tweets linked to YouTube by users labeled as detractors received 0.72 retweets and 0.07 quote tweets on average, with the max number of retweets being 41. While tweets linked to YouTube videos by users labeled as promoters received 14.10 retweets and 1.24 quote tweets on average, with the max number of retweets being 13081. Tweets linked to BitChute videos by users labeled as promoters received 7.56 retweets and 0.66 quote tweets on average, with the max number of retweets being 266. 

\begin{figure*}
    \centering
    \includegraphics[width=0.9\textwidth]{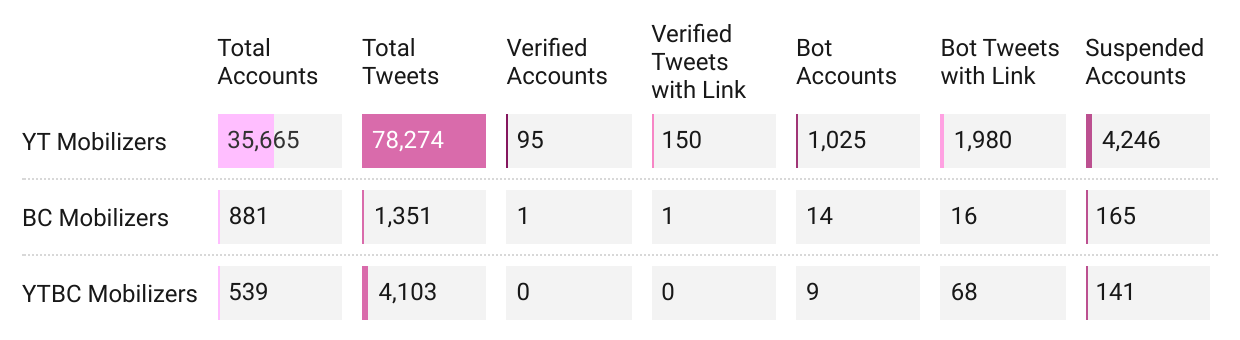}
    \captionof{table}[]{The number of accounts in each mobilizer group that were either verified or bots. The columns are as follows: `Total Accounts' is the total number of accounts in each group. `Total Tweets' is the total number of video-linked tweets produced by each group. `Verified Accounts' is the number of verified accounts in each group. `Verified Tweets with Link' is the number of tweets by verified users in that group which have a link to one of the video platforms. `Bot Accounts' is the number of accounts labeled as a bot by the Botometer API in each group. `Bot Tweets with Link' is the number of tweets by bot accounts in that group which have a link to one of the video platforms. `Suspended accounts' is the number of accounts in each group that were later suspended by Twitter.}
    \label{tbl:bots_veri}
\end{figure*}

\subsubsection{\textbf{How many mobilizers were eventually suspended by Twitter?}} \label{sec:suspend}
As described in \cite{abilov2021voterfraud2020}, after the January 6th U.S. Capitol riots, Twitter suspended accounts that were promoting conspiracy theories related to the election and its outcome. The suspension status of users from this wave of content moderation is another rich piece of metadata provided in the \texttt{VoterFraud2020} dataset. More precisely, the authors labeled users that were later suspended by Twitter in January 2021 (less than 1 month after the original data timeline) \cite{abilov2021voterfraud2020}. As anticipated, the authors note that many of those banned accounts fell into the promoter communities constructed in the dataset. 

This post-collection labeling allows us to \textit{look forward} from the discussion timeline to examine if any cross-platform mobilizers were later suspended on Twitter, providing us with even more insight into the nature of the content being disseminated by these users. Hence, using this metadata, we count the number of suspended accounts in each of our predefined mobilizer groups.  

We found that accounts were suspended in all three groups, but proportionally more were suspended in the BC and YTBC mobilizer groups. As shown in Table \ref{tbl:bots_veri}, we found that 11.9\% of the YT mobilizers were later suspended by Twitter (4246 accounts), while 18.7\% of BC mobilizers (165 accounts) and 26.2\% of YTBC mobilizers (141 accounts) were later suspended by Twitter.

\subsubsection{\textbf{Were mobilizers verified on Twitter or bot accounts?}}
Next, we would like to gain some insight into \textit{who} these mobilizers were. As previous literature on disinformation and media manipulation has noted, disinformation campaigns are often amplified by elite, repeat offenders \cite{bak2021combining}, and the campaign against the 2020 U.S. Presidential Election was no different. For example, we know that elites on Twitter, such as @realDonaldTrump, @LLinWood, @SidneyPowell1, and @GenFlynn, all repeatedly promoted voter fraud claims and were eventually suspended from Twitter \cite{abilov2021voterfraud2020}. At the same time, we know that disinformation messages can also be amplified by bot accounts on Twitter \cite{zannettou2019disinformation, keller2020political}. It is possible that one or both types of accounts (elite accounts and bot accounts) can amplify false claims. In this work, we are examining only a fraction of these message amplifiers: those that linked to external video content.

As a proxy for political elites, we count the number of verified accounts\footnote{Verified accounts on Twitter represent accounts of public interest that are authentic. To receive the blue badge, an account must be ``authentic, notable, and active.'' \url{https://help.twitter.com/en/managing-your-account/about-twitter-verified-accounts}} in each group (YT, BC, or YTBC) using the Twitter API. While not all political elites are verified on Twitter, many are, making this a suitable status indicator. To identify bot accounts in each mobilizer group, we employed the Botometer API \cite{shao2018spread}. In Table \ref{tbl:bots_veri} we show the total number of accounts in each group, number of bots in each group, and number of verified accounts in each group.

First, we found that very few verified accounts shared links to YouTube or BitChute overall: only 0.26\% of accounts that shared a video link were verified. Of those that did, 28 accounts were in the detractor community, including accounts such as @TheYoungTurks and @LateNightSeth. The remaining 68 accounts were in a promoter community, including accounts such as @OANN (One America News Network), @NVGOP (Nevada GOP), @TexasGOP, and multiple Republican candidates for Congress. These verified accounts all shared links to YouTube, with the exception of one account (@minkyungwook, a Korean News Anchor and former politician) who shared a link to a BitChute video entitled: ``DETROIT LEAKS: THIS VIDEO PROVES VOTER FRAUD IN MICHIGAN!!! \#DETROITLEAKS.'' None of these accounts were suspended by Twitter.

Second, we found that bot accounts also played a relatively small role in mobilizing video content, with only 2.8\% of accounts that shared links to video content being labeled as bot accounts. These bot accounts were spread almost evenly across the user groups. Specifically, bot accounts constituted roughly 2.9\% of the total YT mobilizers, 1.6\% of the total YTBC mobilizers, and 1.6\% of the total BC mobilizers. Yet, there is a noticeable difference in the number of bot tweets per group, in particular from the YTBC mobilizers with 7.56 video-linked tweets per bot account. However, with closer examination, this difference in video-linked tweets is all generated by one bot account, which shares a mix of right-wing U.S. politics videos and Christian music videos across both platforms. This bot account appears to maintain very little audience engagement.


\begin{figure*}[ht!]
    \centering
    \includegraphics[width=0.85\textwidth]{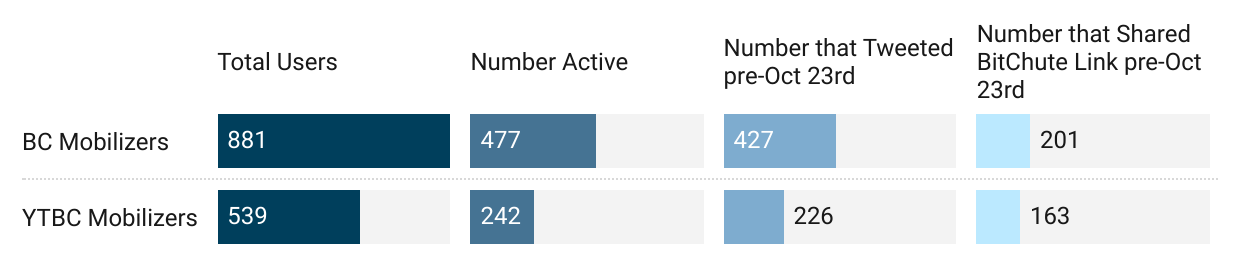}
   \captionof{table}[]{The number of users who shared BitChute links before the VoterFraud2020 dataset. The columns are as follows: `Total' is the total number of users in each group in the \texttt{VoterFraud2020} dataset, `Number Active' is the number of accounts that were still active when our extended dataset was collected, `Number that Tweeted pre-Oct 23rd' is the number of active accounts that had at least one tweet between January 1st, 2020 and October 22nd, 2020, and `Number that Shared BC pre-Oct 23rd' is the number of active users who tweeted at least one BitChute link between January 1st, 2020 and October 22nd, 2020. \label{tbl:bc_before}}
\end{figure*}

\subsubsection{\textbf{Were BC mobilizers sharing BitChute content before the election?}}
The last facet of the mobilizers that we explore is their behavior before the election fraud discussion. Namely, did users who shared BitChute links during the discussion also share links to BitChute previously? Or was the fraud discussion the first time they shared a BitChute link? There is some evidence that major events can lead consumers to new sources of information, such as other social platforms. For example, events with high uncertainty can lead consumers to new sources of information to relieve that uncertainty, or information voids can be maliciously and strategically filled with bad information for political gains \cite{starbird2019disinformation}. Similarly, highly visible political elites may also lead consumers to new information sources, such as the wave of Parler accounts created by members of Congress in early 2021 in response to Twitter's content moderation policies \cite{m2021political}. From this perspective, Twitter users who shared BitChute links may have been led to BitChute as an information source during the discussion of election fraud. 

To this end, we estimate the number of BC mobilizers and YTBC mobilizers who shared BitChute content before the election fraud discussion. We do this by collecting tweets that occurred before the discussion from each active account in the BC and YTBC mobilizer groups, and then examining the links shared. Note, the \texttt{VoterFraud2020} dataset covers tweets between October 23rd, 2020 and December 16th, 2020. Hence, to estimate previous BitChute consumption, we collect all tweets from these users between January 1st, 2020 and October 22nd, 2020 using the Twitter API for Academic Research.

In Table \ref{tbl:bc_before}, we show the results of this link analysis. For each group, roughly half of the accounts were still active when we ran our data collection (447 of 881 in the BC group, 242 of the 539 in the YTBC group), and nearly all of the still active accounts tweeted before October 23rd, 2020 (427 of the 477 in the BC group, 226 of the 242 in the YTBC group). Of these accounts that tweeted before October 23rd, we found that 47.1\% of the BC mobilizers shared at least one tweet with a BitChute link before October 23rd, and 72.1\% of the YTBC mobilizers shared at least one tweet with a BitChute link before October 23rd.

On average, users in the BC mobilizers group tweeted 67.91 tweets with links to BitChute between January 1st and October 22nd, with a maximum number of BitChute linked tweets of 4604 and a minimum of 1. In the YTBC mobilizers group, we see a less skewed distribution of linked tweets per user. On average, users in the YTBC mobilizers group tweeted 83.99 tweets with links to BitChute between January 1st and October 22nd, with a maximum number of BitChute linked tweets of 977 and a minimum of 1.

These results indicate that many of the Twitter users who shared BitChute links promoting election fraud during the 2020 election were already mobilizers of BitChute content on Twitter. Although notably, the most active BitChute link sharers before the election fraud discussion were not necessarily the most active during the discussion. For example, the user who tweeted 4604 BitChute linked tweets before October 23rd, only tweeted one linked tweet after. 

Importantly, these results are likely a lower-bound on the number of users who were actively sharing BitChute links before the election, as nearly half of the BC mobilizers in the \texttt{VoterFraud2020} dataset were no longer active when collecting the historical data.

\subsection{RQ3: How was the content supplied by BitChute different than the content supplied by YouTube?}
While we found very few differences between the Twitter users sharing links to BitChute and YouTube, the content that is being shared from each platform may be very different. This characterization can shed light on a bigger question of the role each platform plays in the information ecosystem. Namely, are the two platforms supplying the same information, distinctly different information, or somewhere in-between? 

To examine this, we build a Structured Topic Model (STM) over the video titles from both platforms, with the platform that the video is from as a binary covariate in the model. STMs are generative models of word counts (similar to other topic modeling methods like LDA \cite{blei2003latent}) that allow for document-level metadata to be used in the model \cite{roberts2019stm}. This additional feature allows us to examine what topics were significantly associated with each platform. The results of this model can be found in Table ~\ref{tbl:topics}. 

\begin{table*}[ht!]
\centering
\fontsize{8.3pt}{9.0pt}
\selectfont
\begin{tabular}{c|c|c|p{7cm}}
\toprule
    \textbf{Topic \#} & \textbf{Stemmed Topic Words} & \textbf{YouTube} & \textbf{Interpretation} \\\midrule
    2 & evid, ballot, giuliani, rudi, wit, reveal, break & +*** & Coverage of Giuliani's lawsuits and interviews with Giuliani, often posted on his own YouTube channel\\\midrule
    10 &  elect, live, hear, state, hold, updat, public & +*** & Live coverage of multiple senate committees and public hearings\\\midrule
    11 & biden, joe, histori, voter, extens, organ, say & +***& Videos about the de-contextualized claim that Joe Biden said he built an ``extensive voter fraud organization''\\\midrule
    6 & powel, sidney, lawsuit, georgia, stop, steal, massiv & +*** &Sidney Powell's claims about Dominion voting machines in the Georgia state election\\\midrule
    7 &  trump, claim, massiv, bombshel, report, uncov, investig & +***& Variety of news coverage of claims made by Trump, channels a mix of mainstream news and alternative news outlets\\\midrule
    3 & elect, video, ntd, arizona, wood, lin, show & +* & Coverage and commentary of Lin Wood and Sidney Powell's lawsuits to overturn state election results, many from the YouTube channel NTD, a self-proclaimed ``global television network founded by Chinese-Americans who fled communism'' \\\midrule
    \midrule
    1 & caught, count, cnn, republican, fact, war, democrat & -*** & Videos claiming voting fraud has been ``caught'' on camera\\\midrule
    4 & barr, america, stopthesteal, state, morn, speech, good & -***& Coverage and analysis of Bill Barr's election fraud comments\\\midrule
    5 & elect, order, leigh, dunda, interfer, execut, detroit & -*** & Commentary on and conspiracies around a 2018 executive order by Trump entitled ``Imposing Certain Sanctions in the Event of Foreign Interference in a United States Election.'' Many of the videos were copies of Facebook videos.\\\midrule
    8 & vote, machin, dominion, hack, elect, watch, ballot & -***& Conspiracy theories about Dominion voting machines being hacked\\\midrule
    9 & fraud, voter, share, kyle, rittenhous, suppress, cold & -*** & Far-right news-like shows covering both voter fraud claims and Kyle Rittenhouse\\\midrule
    12 & vote, michigan, system, live, interview, truth, democrat & -*** & Claims of voter fraud in Michigan, interviews with Dr. Shiva Ayyadurai about voting systems in Michigan\\\midrule
    13 &  expos, presid, news, legal, team, proof, fox & -***& A variety of far-right news-like coverage of Trump's legal team's claims\\
    \bottomrule
\addlinespace[1ex]
\multicolumn{3}{l}{\textsuperscript{***}$p<0.01$, 
  \textsuperscript{**}$p<0.05$, 
  \textsuperscript{*}$p<0.1$}
\end{tabular}
\caption{Topics extracted from video titles using a STM. Note, in the column `YouTube', if there is a $+$, the topic is associated with YouTube. If there is a $-$ the topic is associated with BitChute. Topic interpretation is based on authors qualitatively examining sampled video titles in each topic group. For further discussion on Topic \#11 see \cite{bidenfraud}.}
\label{tbl:topics}
\end{table*}

As expected based on our results in Section \ref{sec:promdetr}, most of the topics across both platforms promoted fraud claims. While the STM indicates significantly different word usage in the video titles from YouTube and BitChute, several of the topics overlap conceptually. For example, videos on both platforms covered the claim that Dominion voting machines were hacked (Topics 6 and 8), claims made by Trump and his legal team (Topics 2, 7, and 13), and claims about fraud in state-level elections (Topics 6, 3, and 12). 

However, there are some notable differences between the videos found in each topic on each platform. On YouTube, multiple topics contained videos of both legitimate news coverage from mainstream news outlets and partisan news coverage from far-right outlets or ``news-like'' YouTube channels. On BitChute, several topics had very specific, niche focuses, such as Topic 5, which was about a 2018 executive order by Trump on foreign interference in a United States Election and how that executive order relates to the 2020 election. At the same time, on BitChute, several topics contained videos that covered a wide variety of conspiracy theories and discussion, such as Topic 9, which contained videos discussing both election fraud claims and the Kyle Rittenhouse shooting case in the same video. On both platforms, many videos had \textit{clickbait-y} titles claiming that proof of fraud during the election was found, although often using different facets of the election as ``proof'' (state-level elections, camera footage from CNN, voting machines, etc.).

These findings indicate that the two platforms are not supplying the same information but also not supplying conceptually different information, rather somewhere in-between. Both covered false claims, conspiracy theories, and hyper-partisan takes on the election. However, YouTube had videos of traditional news coverage from mainstream outlets, and BitChute had videos covering a wider range of conspiracy theories. 

\begin{figure*}[h]
    \centering
    \subfloat[\centering Top channels linked to on BitChute.]{{\includegraphics[width=5.5cm]{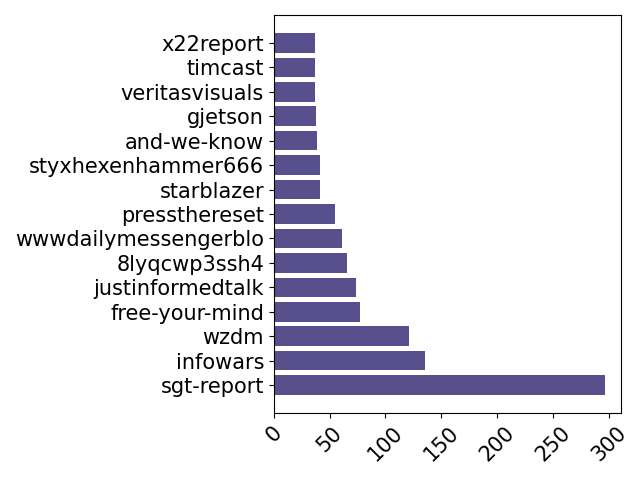}}}%
    \qquad
    \subfloat[\centering Top channels linked to on YouTube.]{{\includegraphics[width=5.5cm]{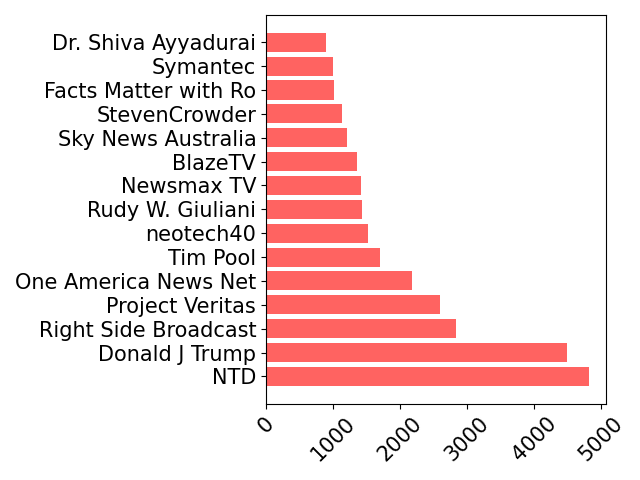}}}\\
    \subfloat[\centering Top channels linked to by the promoter cluster (BitChute or YouTube).]{{\includegraphics[width=5.5cm]{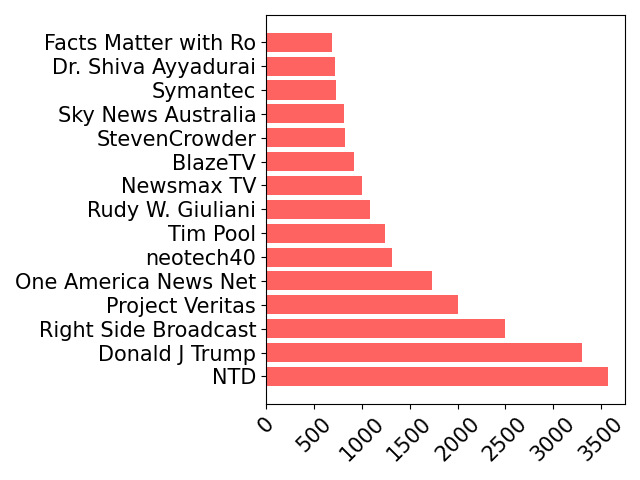}}}%
    \qquad
    \subfloat[\centering Top channels linked to by the detractor cluster (BitChute or YouTube).]{{\includegraphics[width=5.5cm]{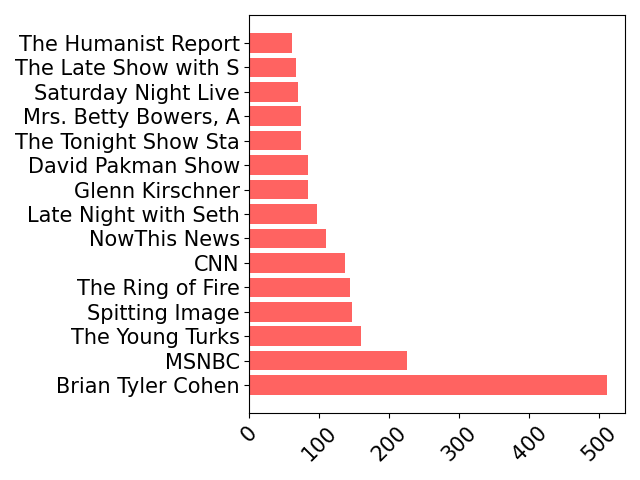}}}%
    \caption{Figures (a) and (b) show the top 15 channels linked to on each platform. Figures (c) and (d) show the top 15 channels linked to by users in the promoter clusters and the detractor cluster. In Figures (c) and (d) we rank across all channels, no matter the platform, yet the top 15 are all YouTube channels. Channel names have been shortened to first 20 characters for visualization. Note the different scales on the x-axes.}%
    \label{fig:top}%
\end{figure*}

\subsection{RQ4: How were the content suppliers on BitChute related to the content suppliers on YouTube?}
One potential reason the topical coverage of videos on both platforms contained some overlap may be that the content producers overlap. Previous work on the relationship between BitChute and YouTube has noted that many prominent content producers maintain accounts on both YouTube and BitChute \cite{trujillo2020bitchute}. To estimate this, we computed the Levenshtein distance (the most common metric for calculating \textit{edit distance}) between channel names on each platform. To aid this analysis, we ranked the top channels linked to in discussion by platform and by user group in Figure \ref{fig:top}.

From this calculation, we found that only 36 of the 5427 total channels had a similarity score of greater than 90\% based on the Levenshtein distance. Three of those 36 matched channels fell into the top 15 most linked-to channels by fraud promoters on Twitter (Project Veritas, StevenCrowder, and Tim Pool). Note that the YouTube versions of their channels were linked to magnitudes more than the BitChute versions.

Additionally, some of these content producers maintain multiple channels on each platform, such as Tim Pool who maintains three channels on each platform: Timcast, TimcastIRL, and TimPool. Again, the YouTube versions of these channels were linked to more than the BitChute versions.

Overall, this approximation suggests that most of the video content is being produced by channel unique to each platform.

\section{Conclusion and Discussion}
This descriptive study has three major takeaways: First, despite YouTube's earlier changes in its content moderation policies, YouTube was still prevalent in the spread of misinformation, disinformation, and conspiracy theories. While BitChute content was linked to and engaged with on Twitter, it played a small role compared to YouTube. This role difference is shown in two ways: First, the number of tweets that both promoted election fraud claims and linked to YouTube videos was much higher than those that linked to BitChute (28 times more); and second, those YouTube linked tweets had more engagement than BitChute linked tweets (14.10 retweets on average versus 7.56 retweets on average).

This finding indicates that there is still more work to do on content moderation policies and practices by `big-tech' platforms. As argued by \citet{wilson2020cross}, given the complexity of cross-platform operations, social media platforms should collaborate on their moderation efforts. The current moderation solutions are often retroactive and siloed, rather than proactive and collaborative. Although, proactive solutions are \textit{much easier said than done}, and more research is needed to implement them. One potential example of collaborative moderation could be: if Twitter detects an emerging disinformation campaign, they could notify YouTube of the campaign to bring it to video moderators' attention, or vice versa. This simple form of collaboration may help control the spread of bad information earlier, rather than relying on suspending accounts well after the campaign. Notably, all platforms in the collaboration would need to be clear and consistent with their policies. 

Second, mobilizers of video content on Twitter, no matter the platform the video content was from, were rarely political elites or bot accounts, but instead appeared to be average Twitter users. Our results suggest amplification of disinformation is participatory. As argued by \citet{starbird2019disinformation}, research on disinformation should move beyond focusing on only bots and trolls to consider the role of online crowds and more complex social structures in the spread and production of disinformation. A key limitation of the evidence found in this paper is that we are not able to determine if any hybrid campaign configurations were behind the spread of cross-platform video content, such as the hybrid configurations described in \cite{jakesch2021trend}, where users voluntarily participated in information campaigns but those campaigns are centrally controlled. Nevertheless, participation of online crowds remains a central point.

Third, while we did not see many well-known, repeat offenders spreading video content on Twitter, we did see well-known, repeat offenders in the linked-to channels on BitChute and YouTube. For example, several of the most linked-to YouTube channels are channels that have produced disinformation well before the election fraud discussions (e.g. One America News Network, NewsmaxTV). Similarly, several of the most linked-to BitChute channels are channels known for producing conspiracy theories (e.g. infowars, x22report). 

Past research has shown that banning or moderating users who repeatedly produce disinformation and are highly followed can be a useful approach \cite{bak2021combining}. A practical example of this comes from Zignal Labs, who showed a 73\% decline in election fraud discussion after Donald Trump was suspended from Twitter\footnote{\url{https://www.washingtonpost.com/technology/2021/01/16/misinformation-trump-twitter/}}. These past results paired with the results in this paper suggest that it may be effective for Twitter to automatically moderate tweets with URLs to these known, repeat offenders from other platforms. 

Ultimately, these results suggest the roles of elites -- both political elites and elite platforms, like YouTube -- remain as core sources of mis- and disinformation in online spaces. That said, less mainstream and influential entities, such as alt-tech platforms and common information consumers, remain a key participatory element in the spread of these disinformation campaigns. Efforts to counter such campaigns must improve, both in mainstream platforms’ willingness to collaborate and in how they respond to and moderate the fringe elements (both alt-tech and extreme consumers) in this space.

\bibliographystyle{ACM-Reference-Format}
\bibliography{references}
\end{document}